\input harvmac
\noblackbox
\newcount\figno
\figno=0
\def\fig#1#2#3{
\par\begingroup\parindent=0pt\leftskip=1cm\rightskip=1cm\parindent=0pt
\baselineskip=11pt \global\advance\figno by 1 \midinsert
\epsfxsize=#3 \centerline{\epsfbox{#2}} \vskip 12pt {\bf Fig.
\the\figno:} #1\par
\endinsert\endgroup\par
}
\def\figlabel#1{\xdef#1{\the\figno}}
\def\encadremath#1{\vbox{\hrule\hbox{\vrule\kern8pt\vbox{\kern8pt
\hbox{$\displaystyle #1$}\kern8pt} \kern8pt\vrule}\hrule}}

\input epsf

\overfullrule=0pt

%
\def\tilde{\widetilde}
\def\bar{\overline}
\def\np#1#2#3{Nucl. Phys. {\bf B#1} (#2) #3}

%
\def\inbar{\,\vrule height1.5ex width.4pt depth0pt}
\def\IB{\relax{\rm I\kern-.18em B}}
\def\IC{\relax\hbox{$\inbar\kern-.3em{\rm C}$}}
\def\ID{\relax{\rm I\kern-.18em D}}
\def\IE{\relax{\rm I\kern-.18em E}}
\def\IF{\relax{\rm I\kern-.18em F}}
\def\IG{\relax\hbox{$\inbar\kern-.3em{\rm G}$}}
\def\IH{\relax{\rm I\kern-.18em H}}
\def\II{\relax{\rm I\kern-.18em I}}
\def\IK{\relax{\rm I\kern-.18em K}}
\def\IL{\relax{\rm I\kern-.18em L}}
\def\IM{\relax{\rm I\kern-.18em M}}
\def\IN{\relax{\rm I\kern-.18em N}}
\def\IO{\relax\hbox{$\inbar\kern-.3em{\rm O}$}}
\def\IP{\relax{\rm I\kern-.18em P}}
\def\IQ{\relax\hbox{$\inbar\kern-.3em{\rm Q}$}}
\def\IR{\relax{\rm I\kern-.18em R}}
\font\cmss=cmss10 \font\cmsss=cmss10 at 7pt
\def\IZ{\relax\ifmmode\mathchoice
{\hbox{\cmss Z\kern-.4em Z}}{\hbox{\cmss Z\kern-.4em Z}}
{\lower.9pt\hbox{\cmsss Z\kern-.4em Z}} {\lower1.2pt\hbox{\cmsss
Z\kern-.4em Z}}\else{\cmss Z\kern-.4em Z}\fi}
\def\IGa{\relax\hbox{${\rm I}\kern-.18em\Gamma$}}
\def\IPi{\relax\hbox{${\rm I}\kern-.18em\Pi$}}
\def\ITh{\relax\hbox{$\inbar\kern-.3em\Theta$}}
\def\IOm{\relax\hbox{$\inbar\kern-3.00pt\Omega$}}

\def\p{\partial}

\font\zfont = cmss10 
 
\def\bigone{\hbox{1\kern -.23em {\rm l}}}
\def\ZZ{\hbox{\zfont Z\kern-.4emZ}}

\def\e{\epsilon}

\def\th{\theta}

\def\p{\phi}

\def\IR{\relax{\rm I\kern-.18em R}}
\def\I1{\relax{\rm I\kern-.6em 1}}
\def\Dsl{\,\raise.15ex\hbox{/}\mkern-13.5mu D}
\def\Gsl{\,\raise.15ex\hbox{/}\mkern-13.5mu G}
\def\Csl{\,\raise.15ex\hbox{/}\mkern-13.5mu C}
\font\cmss=cmss10 \font\cmsss=cmss10 at 7pt

\font\zfont = cmss10 
 
\def\bigone{\hbox{1\kern -.23em {\rm l}}}
\def\ZZ{\hbox{\zfont Z\kern-.4emZ}}

\def\unlockat{\catcode`\@=11}
\def\lockat{\catcode`\@=12}

\unlockat

\def\newsec#1{\global\advance\secno by1\message{(\the\secno. #1)}
\global\subsecno=0\global\subsubsecno=0\eqnres@t\noindent
{\bf\the\secno. #1} \writetoca{{\secsym}
{#1}}\par\nobreak\medskip\nobreak}
\global\newcount\subsecno \global\subsecno=0
\def\subsec#1{\global\advance\subsecno  
by1\message{(\secsym\the\subsecno. #1)}
\ifnum\lastpenalty>9000\else\bigbreak\fi\global\subsubsecno=0
\noindent{\it\secsym\the\subsecno. #1} \writetoca{\string\quad
{\secsym\the\subsecno.} {#1}}
\par\nobreak\medskip\nobreak}
\global\newcount\subsubsecno \global\subsubsecno=0
\def\subsubsec#1{\global\advance\subsubsecno by1
\message{(\secsym\the\subsecno.\the\subsubsecno. #1)}
\ifnum\lastpenalty>9000\else\bigbreak\fi
\noindent\quad{\secsym\the\subsecno.\the\subsubsecno.}{#1}
\writetoca{\string\qquad{\secsym\the\subsecno.\the\subsubsecno.}{#1}}
\par\nobreak\medskip\nobreak}
\def\subsubseclab#1{\DefWarn#1\xdef
#1{\noexpand\hyperref{}{subsubsection}%
{\secsym\the\subsecno.\the\subsubsecno}%
{\secsym\the\subsecno.\the\subsubsecno}}%
\writedef{#1\leftbracket#1}\wrlabeL{#1=#1}}
\lockat

\def\IP{\relax{\rm I\kern-.18em P}}
\def\IC{\relax{\rm I \kern-.5em C}}
\def\IZ{\relax{\rm I\kern-.18em Z}}
\lref\np{B.~E.~W.~Nilsson, C.~N.~Pope,
{\sl Hopf fibration of eleven-dimensional supergravity},
 Class.Quant.Grav. {\bf 1}, 499 (1984). } 
\lref\dhis{M.~J.~Duff,
P.~S.~Howe, T.~Inami and K.~S.~Stelle, {\sl Superstrings In D = 10 From
Supermembranes In D = 11}, Phys. Lett. B {\bf 191}, 70 (1987).} 
\lref\dlp{M.~J.~Duff, H.~Lu and C.~N.~Pope, {\sl Supersymmetry without
supersymmetry}, Phys.\ Lett.\ B {\bf 409}, 136 (1997) [hep-th/9704186].  
} 
\lref\dlpun{M.~J.~Duff, H.~Lu and
C.~N.~Pope, {\sl AdS(5) x S(5) untwisted}, Nucl.\ Phys.\ B {\bf 532}, 181
(1998) [hep-th/9803061]. 
} 
\lref\kt{I.~R.~Klebanov and A.~A.~Tseytlin, {\sl Gravity duals of
supersymmetric SU(N) x SU(N+M) gauge theories}, Nucl.\ Phys.\ B {\bf 578},
123 (2000) [hep-th/0002159].  
} 
\lref\ks{I.~R.~Klebanov and M.~J.~Strassler, {\sl Supergravity and a
confining gauge theory: Duality cascades and chiSB-resolution of naked
singularities}, JHEP {\bf 0008}, 052 (2000) [hep-th/0007191].  
} 
\lref\cdlo{P.~Candelas, X.~C.~de la Ossa,
{\sl Comments on Conifolds}, Nucl.\ Phys.\ {\bf B342}, 246 (1990).} 
\lref\amv{M.~Atiyah, J.~Maldacena and C.~Vafa, {\sl An M-theory flop as a
large n duality} [hep-th/0011256].  
} 
\lref\vafan{C.~Vafa, {\sl Superstrings and topological strings at large
N} [hep-th/0008142].  
} 
\lref\mt{R.~Minasian and D.~Tsimpis, {\sl On the geometry of non-trivially
embedded branes}, Nucl.\ Phys.\ B {\bf 572}, 499 (2000) [hep-th/9911042].  
} 
\lref\gk{S.~S.~Gubser and I.~R.~Klebanov, {\sl
Baryons and domain walls in an N = 1 superconformal gauge theory}, Phys.\
Rev.\ D {\bf 58}, 125025 (1998) [hep-th/9808075].  
} 
\lref\dmb{K.~Dasgupta and S.~Mukhi, {\sl Brane
constructions, fractional branes and anti-de Sitter domain walls}, JHEP
{\bf 9907}, 008 (1999) [hep-th/9904131].  
}
\lref\kn{I.~R.~Klebanov and N.~A.~Nekrasov, {\sl Gravity duals of
fractional branes and logarithmic RG flow}, Nucl.\ Phys.\ B {\bf 574}, 263
(2000) [hep-th/9911096].  
} 
\lref\dnp{M.~J.~Duff, B.~E.~W.~Nilsson, C.~N.~Pope, {\sl Kaluza-Klein
supergravity}, Phys.\ Rep.\ {\bf 130}, 1 (1986).} 
\lref\dlpb{M.~J.~Duff, H.~Lu and C.~N.~Pope, {\sl AdS(3) x S**3
(un)twisted and squashed, and an O(2,2,Z) multiplet of dyonic strings},
Nucl.\ Phys.\ B {\bf 544}, 145 (1999) [hep-th/9807173].  
} 
\lref\wia{E.~Witten, {\sl Solutions of four-dimensional field theories via
M-theory}, Nucl.\ Phys.\ B {\bf 500}, 3 (1997) [hep-th/9703166].
} 
\lref\wib{E.~Witten, {\sl Branes and
the dynamics of {QCD}}, Nucl.\ Phys.\ B {\bf 507}, 658 (1997)
[hep-th/9706109].  
} 
\lref\dma{K.~Dasgupta and S.~Mukhi, {\sl Brane constructions, conifolds
and M-theory}, Nucl.\ Phys.\ B {\bf 551}, 204 (1999) [hep-th/9811139].  
} 
\lref\clpc{M.~Cvetic, H.~Lu and
C.~N.~Pope, {\sl Decoupling limit, lens spaces and Taub-NUT: D = 4 black
hole microscopics from D = 5 black holes}, Nucl.\ Phys.\ B {\bf 549}, 194
(1999) [hep-th/9811107]. 
} 
\lref\ghm{R.~Gregory, J.~A.~Harvey and G.~Moore, {\sl Unwinding strings
and T-duality of Kaluza-Klein and H-monopoles}, Adv.\ Theor.\ Math.\
Phys.\ {\bf 1}, 283 (1997) [hep-th/9708086].  
} 
\lref\zt{L.~A.~Pando Zayas and A.~A.~Tseytlin, {\sl
3-branes on resolved conifold}, JHEP {\bf 0011}, 028 (2000)
[hep-th/0010088]. 
} 
\lref\cglpa{M.~Cvetic, H.~Lu and C.~N.~Pope, {\sl Massless 3-branes in
M-theory} [hep-th/0105096].  
} 
\lref\cglpb{M.~Cvetic, G.~W.~Gibbons, H.~Lu and C.~N.~Pope, {\sl
M3-branes, G(2) manifolds and pseudo-supersymmetry} [hep-th/0106026].
} 
\lref\SAcharya{B.~S.~Acharya, {\sl On realising N = 1 super Yang-Mills in
M theory} [hep-th/0011089]. 
} 
\lref\DEdelsteinNunez{J.~D.~Edelstein and C.~Nunez, {\sl D6 branes and
M-theory geometrical transitions from gauged supergravity}, JHEP {\bf
0104}, 028 (2001) [hep-th/0103167]. 
}
\lref\DasguptaOhTatar{K.~Dasgupta, K.~Oh and R.~Tatar, {\sl Geometric
transition, large N dualities and MQCD dynamics} [hep-th/0105066].
} 
\lref\AganagicVafa{M.~Aganagic and C.~Vafa, {\sl Mirror symmetry and a
G(2) flop} [hep-th/0105225]. 
}
\lref\BrandhuberGomisSGubserGukov{A.~Brandhuber,
J.~Gomis, S.~S.~Gubser and S.~Gukov, {\sl Gauge theory at large N and new
G(2) holonomy metrics} [hep-th/0106034]. 
}
\lref\mnz{
J.~M.~Maldacena and C.~Nunez,
{\sl Towards the large N limit of pure N = 1 super Yang Mills},
Phys.\ Rev.\ Lett.\  {\bf 86}, 588 (2001)
[hep-th/0008001].
}

%
\Title{\vbox{\baselineskip12pt \hbox{CPHT-S032.0601}
\hbox{G{\"o}teborg ITP preprint}
\hbox{hep-th/0106266} }} 
{\vbox{ \centerline{Hopf reductions, fluxes and branes}
}}

\bigskip
\centerline{Ruben Minasian,$^{\dagger}$ and Dimitrios
Tsimpis$^{\ddagger}$}

\bigskip
\centerline{$^\dagger${\sl  Centre de Physique Th{\'e}orique, Ecole
Polytechnique}\footnote{${}^\circ$}{\sevenrm Unit{\'e} mixte du
CNRS et de l'EP, UMR 7644.}}
\centerline{ \sl F-91128 Palaiseau, France}
\centerline{\tt ruben@cpht.polytechnique.fr}
\medskip

\centerline{$^\ddagger$ {\sl Department of Theoretical Physics}}
\centerline{\sl G{\"o}teborg University and Chalmers University of
Technology}
\centerline{\sl SE-412 96 G{\"o}teborg, Sweden}
\centerline{\tt tsimpis@fy.chalmers.se}

\bigskip
\centerline{\bf Abstract}
\medskip

\noindent We use a series of reductions, $T$-dualities and liftings to
construct connections between fractional
brane solutions in IIA, IIB and M-theory. We find a number of
phantom branes that are not supported by the geometry, however materialize
upon untwisting and/or Hopf-reduction.

\Date{June 2001}

\vfill\eject

\newsec{Introduction and discussion}

The study of M-theory backgrounds with $U(1)$ 
isometries and their associated
Hopf reductions dates back to the eighties (see e.g. \refs{\np, \dhis}). 
In the presence of branes and depending on whether 
the direction of the reduction is parallel or transverse to
the brane, the dimensionality
of the brane worldvolume may or may not change under the reduction. 
A subsequent $T$-duality transformation can be applied. 
In the case of $AdS_{D-d} \times S^{d}$ backgrounds the fact that 
both the $AdS_{D-d}$ and the $ S^{d}$ factors can be thought of
as $U(1)$ fibrations has been used extensively in order
to establish new vacua through Hopf-reductions/$T$-dualities.
One of the lessons was that these operations can untwist the 
$U(1)$ bundles and break supersymmetry at the level of supergravity
solutions (although this may not be true at the level of
full string theory) \refs{\dlp, \dlpun, \dlpb}. 
The recent interest in supergravity solutions involving branes
wrapping cycles in conifold backgrounds motivates us to
investigate some reductions and dualities in this case. 

Our starting point is the singular type IIB solution of \kt, (the KT
solution). 
The geometry is of the form 
(warped) $\IR^{1,3}\times {\cal C}(T^{1,1})$ where 
the transverse space ${ \cal C}(T^{1,1})$ is the
conifold of \cdlo, which is a cone over $T^{1,1}$.  
The latter is topologically $S^2\times S^3$.
 Since this can be thought of as a
$U(1)$ fibration over $S^2 \times S^2$, 
we can Hopf-reduce/$T$-dualize along the $U(1)$. The KT solution
involves a system of ordinary and fractional D3 branes. The latter 
are thought of in this context as D5 branes wrapping the $S^2$
of  $T^{1,1}$. The $T$-duality in this case will be along
a direction transverse to the branes.

A closely related non-singular system is the KS solution \ks. 
In this case the transverse space is instead the deformed conifold
of \cdlo. The latter is obtained from the conifold by replacing the apex
by an $S^3$. In the context of KS, the D5 wrapping the $S^2$ produces flux
through the $S^3$, which is thus stabilized (kept finite). The deformed
conifold is also a $U(1)$ fibration,  although  Hopf-reducing is more
complicated technically.

In a related development \refs{\SAcharya,\amv}, it was realized that a
system of D6 branes
in IIA theory wrapping the $S^3$ of the resolved conifold 
(the AMV solution) can be
lifted in M-theory to a manifold of $G_2$ holonomy. 
The resolved conifold is obtained from the conifold by replacing the apex 
by an $S^2$. In the case of AMV, the D6-branes wrapping the $S^3$ produce
flux through the $S^2$, which is stabilized.\foot{In the terminology  of
\amv, this is the situation
where the D6 branes have disappeared and have been replaced
by the $F_2$ flux. Here we use the notion of ``brane'' modulo the large $N$
duality of \vafan, namely both for $N$ D6-branes wrapping $T^*S^3$ and for
the physically equivalent case of the resolved conifold with $N$ units of
flux through the $S^2$.}
Once more,
metamorphoses of $U(1)$ bundles in  this and related setups provide 
interesting connections.  

This paper contains a series of Hopf-reductions, $T$-dualities and liftings
involving branes in the geometries discussed above.  
There are two points that
we find worth emphasizing:

While $T$-duality can untwist the $U(1)$ bundles, the M-theory lifting (in
the presence of fractional branes) can induce new twistings. This has
received much attention lately due to \amv. In the liftings
considered here, orientation
issues play a subtle role. 
Starting from IIB theories in a background involving a $T^{1,1}$
space we may 
end up in an M-theory background with a different $T^{1,1}$.

We find a number of examples with twisted  geometries 
(with respect to the $U(1)$ fibre) where 
there are fluxes, but 
where there are no corresponding non-contractible cycles 
for the fluxes to be
integrated over into charges. 
We think of these backgrounds as phantom branes that are not
supported by the (twisted) geometry. 
However, there are two instances in which real branes can emerge 
from phantom ones. $a$) By switching-off some fluxes
(typically corresponding to fractional branes) the geometry may
get untwisted giving thus rise to the cycles necessary for supporting
the branes. Note that this is not a continuous process since
it involves taking the number of fractional branes 
(an integer) to zero. $b$) By Hopf-reduction.

In a way this could be considered the discrete
analogue of the large-$N$ duality between fluxes in twisted geometry 
and branes. A somewhat similar phenomenon
which has already appeared 
in the literature is the case of winding strings in
the background of the Kaluza-Klein monopole \ghm. 
We comment on this in section 3.

The outline of the paper is as follows. Since the properties of 
$T^{1,1}$ spaces will be important in the following, we
review these in section 2.  Special attention is paid to issues
related to orientations of cycles. The $T$-dual and the
M-theory
lift of  the KT solution  and its $S$-dual are discussed in sections 3
and 4 respectively. Here  we find a number of phantom branes that
are not supported by the geometry, however materialize upon 
Hopf-reduction and/or 
untwisting. Section 5 is more speculative and 
concerns fivebranes in the geometry of AMV. 
Finally in section 6 we explore the $U(1)$ isometry of the
deformed conifold and find  a new set of variables in which this
isometry is explicit. Some useful formulae concerning different reductions
and $T$-dualities are collected in the appendix.

\newsec{The geometry of $T^{p,q}$ spaces}

$T^{p,q}$ spaces  can either be thought of as
$U(1)$ fibrations over $S^2 \times S^2$
or as $SU(2)\times SU(2)/U(1)$ coset spaces. Let $0\leq \phi_i \leq 2\pi, 
\,\, 0\leq \theta_i \leq \pi , \,\, i=1,2 $ parametrize the
 two $S^2$ and let $0\leq \psi \leq 4\pi$ be the coordinate of the $U(1)$ fibre.
The most general $T^{p,q}$ metric reads  
\eqn\ta{\eqalign{
ds_T^2 =D&(d\psi+p\, cos\theta_1 d\phi_1
+q\, cos\theta_2 d\phi_2)^2+A(sin^2\theta_1 d\phi_1^2 + d\theta_1^2)\cr
&+C(sin^2\theta_2 d\phi_2^2 + d\theta_2^2) 
+2B[cos\psi (d\theta_1 d\theta_2- sin\theta_1 sin\theta_2 d\phi_1 d\phi_2)\cr
&+sin\psi (sin\theta_1 d\phi_1 d\theta_2+sin\theta_2 d\phi_2 d\theta_1)]
\cr}
}
where $A,B,C,D$ are constants.

It is convenient to introduce the following basis 
of one-forms, motivated by the geometry of coset spaces \mt:
%
%
%
\eqn\baseb{\eqalign{ \pmatrix{ e^1 \cr 
e^2 \cr }&= \pmatrix{ sin\theta_1 d\phi_1 \cr 
d\theta_1 \cr } \cr
\pmatrix{ e^3 \cr 
e^4 \cr }&= \pmatrix{ cos\psi &
-sin\psi \cr  sin\psi &  cos\psi \cr }
\pmatrix{  sin\theta_2 d\phi_2 \cr 
d\theta_2 \cr } \cr
e^5=d\psi+A^{(p,q)}_1&; \,\,\,\,\,\,\, 
A^{(p,q)}_1=p\, cos\theta_1 d\phi_1+q\, cos\theta_2 d\phi_2.
\cr}}
$A^{(p,q)}_1$ is the connection 1-form of the $(p,\, q)$ $U(1)$ bundle
over $S^2\times S^2$.
In terms of the above base the metric \ta\ becomes
\eqn\tb{
ds_T^2=D(e^5)^2+A\left((e^1)^2+(e^2)^2\right)+
C\left((e^3)^2+(e^4)^2\right)+2B(e^1 e^3+e^2 e^4)
}
Of particular interest to us is the space $T^{1,1}$ 
which is topologically $S^3\times S^2$. A basis
for the harmonic representatives of the (one-dimensional)
spaces $ H^2(T^{1,1}, \ZZ), \,  H^3(T^{1,1}, \ZZ)$ was 
constructed in \gk:
\eqn\ktde{
\eqalign{
\omega_2&= (sin\theta_1   d\phi_1 \wedge d\theta_1 -sin\theta_2 
 d\phi_2 \wedge d\theta_2)\cr
&=e^1\wedge e^2-e^3\wedge e^4\cr
\omega_3&=e^5 \wedge \omega_2;\cr
}}
Both $\omega_2$ and $\omega_3$ are closed. Locally we can write
\eqn\defo{\omega_2=-d\omega_1; \,\,\,\,\,\,\, 
\omega_1= cos\theta_1 d\phi_1-cos\theta_2 d\phi_2}
We denote by $S^2, \, S^3$ the basis of homology cycles for the 
spaces $H_2(T^{1,1},\ZZ), \, H_3 (T^{1,1},\ZZ)$ so that
\eqn\homc{\int_{S^3} \omega_3=\int_{S^2} \omega_2=1}
Since $b_2(T^{1,1})=1 \, (b_3(T^{1,1})=1)$, there is only one $S^2$ 
($S^3$) homologically. Consider 
the two different 3-spheres corresponding to the orbits of each
of the two $SU(2)$
factors in the isometries of $T^{1,1}$. Let us  denote 
them by $\Sigma^{(i)}_3\, , i=1,2$. They are parametrized
by $(\phi_i,\theta_i,\psi)$. Similarly, 
let us denote by $\Sigma^{(i)}_2$ the two 2-spheres 
parametrized by 
$(\phi_i,\theta_i)\, ,i=1,2$.
The homology basis can be written as \refs{\gk, \dmb}:
\eqn\poiu{S^3=\Sigma^{(1)}_3-\Sigma^{(2)}_3; \,\,\,\,\,\,\,
S^2=\Sigma^{(1)}_2-\Sigma^{(2)}_2.}
Following \refs{\gk, \dmb} we will interpret a (fractional) 
brane wrapping $\Sigma^{(1)}$ as being equivalent to an antibrane
wrapping $\Sigma^{(2)}$.
Under the transformation 
$\theta_2\leftrightarrow \pi-\theta_2$,  $\omega_1$
becomes the $U(1)$ connection over a $T^{1,1}$
with {\it reversed} orientation. 
We have, 
\eqn\ortr{\eqalign{
A_1&\leftrightarrow \omega_1\cr
\omega_2&\leftrightarrow  dA_1\cr
\Sigma^{(2)}_{2,3}&\leftrightarrow -{\Sigma}^{(2)}_{2,3}.
\cr}}
where $A_1:=A^{(1,1)}_1$
Under this orientation-reversal, 
a brane wrapping $\Sigma^{(2)}$ becomes an antibrane and vice versa.
We also note that \homc\ can be written as
\eqn\cyci{\int_{\Sigma^{(1)}}\omega-\int_{\Sigma^{(2)}}\omega=1.}
On the other hand,
\eqn\cyci{\int_{\Sigma^{(1)}}\omega+\int_{\Sigma^{(2)}}\omega=0.}
Taking \ortr\ into account, 
a similar set of relations can be easily derived for 
the integrals of $dA_1, \, e^5\wedge dA_1$ over $\Sigma_{2,3}$.

Finally we can define the $T^{1,-1}$ space
which differs from $T^{1,1}$ in the relative sign
between the two factors in the $U(1)$ connection or, equivalently, 
in the relative orientation of the two $S^2$ in the base.
An equation similar to \ortr\ relates 
the  $T^{1,-1}$ with its orientation-reversal. Note that 
$A_1$ can be either the $U(1)$ connection of a $T^{1,1}$ or of a 
$T^{1,-1}$ with reversed orientation. Similarly, $\omega_1$
can be either the $U(1)$ connection of a $T^{1,-1}$ or of a 
$T^{1,1}$ with reversed orientation.

\newsec{The Klebanov-Tseytlin solution.}

We start from the ten-dimensional IIB singular solution for $N$
ordinary and $M$ fractional D3 branes, presented in \kt.
The geometry is of the form (warped) 
$R^{1,3}\times {\cal C}(T^{1,1})$ where 
the transverse space ${ \cal C}(T^{1,1})$ is the
conifold of \cdlo\ which is a cone over $T^{1,1}$.  
$T^{1,1}$ has a $U(1)$ isometry and we can apply the techniques of
Hopf-reduction/$T$-duality
explained in the appendix. Reducing along the $U(1)$ to nine-dimensional IIB, 
dualizing to  IIA and lifting back to ten dimensions, we find that the
$U(1)$ bundle  gets ``untwisted'' so that the transverse space is foliated
by  $\Sigma^{(1)}_2\times \Sigma^{(2)}_2 \times S^1$. 
We will see how in a further lifting to eleven-dimensions a new
``twisted''
direction develops so that the transverse space is foliated by
$T^{1,1}\times S^1$.

The KT  solution reads
\eqn\ktso{
\eqalign{
e^{\phi}=g_s; \,\,\,\,\,\,\, B_2=3g_s M\, ln{r\over r_0}\, \omega_2;
\,\,\,\,\,\,\, F_3=M \omega_3 \cr
F_5={\cal F}_5+{}^* {\cal F}_5;  \,\,\,\,\,\,\, 
{\cal F}_5=(N+3g_s M^2ln{r\over r_0})\, \omega_2 \wedge \omega_3 \cr}
}
where $\omega_2, \, \omega_3$ where defined in the previous section.
The metric (in the string frame) is
\eqn\ktme{
ds_{10}^2=h^{-{1 \over 2}}dx^{\mu} dx_{\mu}+ h^{1\over 2}
(dr^2+r^2 ds_T^2)
}
where
\eqn\ktcf{
h(r)= b_0+ {c \times g_s \over r^4}(N+{3\over 4}g_s 
M^2+3g_s M^2 ln{r\over r_0}  )
}
The metric $ds^2_T$ is the $T^{1,1}$ metric defined in 
the previous section for the special case $D=1/9, \, A=1/6, \, B=0$.
With this choice of constants $T^{1,1}$ becomes Einstein. 
The constant $b_0$ can be fixed by demanding that the asymptotically
 flat limit of the metric be the standard Minkowski. Similarly 
the constant $c$ can be fixed by requiring that for $M=0$ the metric 
reduce 
to the standard $AdS_5 \times S^5 $ in the 
near-horizon limit. We also have \foot{We are droping
numerical constants of proportionality.},
\eqn\ktfq{
\int_{S^3} F_3 = M; \, \,\,\,\,\,\, \int_{T^{1,1}} F_5 
= N +3g_s M^2 ln{r\over r_0}
}
The identification of fractional D3 branes with D5 branes wrapping
the $S^2$ of the conifold, was justified in \kn.

\subsec{The IIA versions of KT solution}

As already mentioned, reducing the KT solution along the $U(1)$ fibre to
nine-dimensional IIB, dualizing to IIA and lifting to ten-dimensional 
IIA, we find that the $U(1)$
bundle gets ``untwisted'' so that the transverse space is foliated by
$\Sigma^{(1)}_2\times \Sigma^{(2)}_2 \times S^1$.  The interpretation of
the $T$-dualized (IIA) version of KT, depends on our choice of orientation
for $\Sigma^{(2)}$. Geometrically speaking, there seems to be no reason
for preferring one orientation over the other, but the brane content is
entirely different in each case.

It must be noted  that while reversing the orientation of a solution is
also a solution to the supergravity equations, the amount of preserved
supersymmetry  may
change. In fact, if an Einstein space that is not a
round  sphere has Killing spinors, its orientation reversal gives a space
with  no Killing spinors \dnp. So trying to preserve the original amount of  
supersymmetry will choose an orientation for us, however we will 
consider both cases here. Concerning 
the effect of $T$-duality/Hopf-reduction we note
that 
$T$-duality at the full string theory level
always preserves as much supersymmetry as has survived the $U(1)$
compactification. The $U(1)$ compactification in the case where
the compactifying direction is naturally periodic (as is here)
also leaves supersymmetry unbroken at the full string theory level.
At the level of supergravity however, it is known that these
operations  can brake the supersymmetries of p-brane solutions 
(see \dlpb\ for a detailed discussion).

\subsubsec{ T-duality}

 Let us start with the orientation for $\Sigma^{(2)}_2$ 
which is inherited from \poiu. Reducing to nine dimensions, dualizing and
lifting back to ten-dimensional type IIA, we obtain the  following
solution
\eqn\ktds{
\eqalign{
F_2=M \omega_2&;  \,\,\,\,\,\,\, 
F_3= 3g_s M\, {dr \over r}\wedge \omega_2+ dA_1 \wedge dz \cr
&F_4=(N+3g_s  M^2 ln{r\over r_0})\, \omega_2 
\wedge \omega_2 \cr
}
}
while the metric (in the Einstein frame) reads
\eqn\ktdm{
g_s^{1 \over 2}ds_{10}^2= ({1\over 9} h^{1\over 2}r^2 )^{1\over 4}
\left( h^{-{1 \over 2}}dx^{\mu} dx_{\mu}+ h^{1\over 2}
(dr^2+{1\over 6} r^2 \sum_{i=1}^2 (d\Sigma_2^{(i)})^2)\right) 
+({1\over 9} h^{1\over 2}r^2 )^{-{3\over 4}} dz^2
}
where $ (d\Sigma_2^{(i)})^2:=sin^2\theta_i d\phi^2_i +d\theta_i^2$.
We see that the dualization has untwisted the $U(1)$ bundle so
that the transverse space is foliated by $S^2\times S^2 \times S^1$.

There is  a charge $Q^{(6)}$ 
corresponding to D6 branes wrapping an $S^2\times S^1$.
This gives fractional D3 branes! Indeed, let's denote by $\Sigma_{1,2}$
the two $S^2$ factors. We have
\eqn\dsa{  Q^{(6)}= \int_{\Sigma^{(1)}_2} F_2-
\int_{\Sigma^{(2)}_2} F_2 = M  }
as can be seen from \homc.

There is however zero net $Q^{(5)}$ charge corresponding
to NS5 branes wrapping an $S^2$,
\eqn\dfa{
Q^{(5)}=\int_{\Sigma^{(1)}_2\times S^1} F_3
-\int_{\Sigma^{(2)}_2\times S^1} F_3=0
}
Finally, there is a $Q^{(4)}$ charge corresponding to D4 branes
wrapping $S^1$
\eqn\dfoa{
Q^{(4)}=\int_{\Sigma^{(1)}_2\times \Sigma^{(2)}_2} F_4 
= N+3g_s M^2 ln{r\over r_0}
}
The brane-content is in agreement with the (naive) 
expectations from $T$-duality:
The D5 (fractional D3) become D6, the D3 become D4 wrapping
the $T$-duality circle.

\subsubsec{ The orientation-reversed version}
We now perform an orientation-reversal transformation on the KT solution 
in IIA. Upon changing the orientation, the geometrical 
interpretation of the solution changes. This is because a brane wrapping 
a nontrivial cycle becomes an antibrane upon changing the 
orientation of the cycle. More specifically
there is now zero net  charge $Q^{(6)}$ 
corresponding to D6 branes wrapping an $S^2\times S^1$
\eqn\dsab{  Q^{(6)}= \int_{\Sigma^{(1)}_2} F_2+ 
\int_{\Sigma^{(2)}_2} F_2 =0  }
In other words, 
there is an equal number of D6 branes and antibranes.

There is however  nonzero net $Q^{(5)}$ charge corresponding
to an NS5 brane wrapping $S^2$,
\eqn\dfab{
Q^{(5)}=\int_{\Sigma^{(1)}_2\times S^1} F_3
+\int_{\Sigma^{(2)}_2\times S^1} F_3 = 1
}

The $Q^{(4)}$ charge corresponding to D4 branes
wrapping $S^1$ is still given by
\eqn\dfoatut{
Q^{(4)}=\int_{\Sigma^{(1)}_2\times \Sigma^{(2)}_2} F_4 
= N+3g_s M^2 ln{r\over r_0}
}
The above brane content should be compared with the system 
of \refs{\wia, \wib, \dma, \dmb} where there are two  NS5 
branes, one of them stretching along the directions $x^{0-5}$, 
and the other along $x^{0-3,7,8}$. 
The $x^6$ direction is a circle. 
There are also two types of D4 branes along $x^{0-3,6}$. 
$N$ of them going around the circle
and $M$ of them stretching between the two NS5 branes.

In order to compare to the situation at hand, 
we should identify $x^{4,5}$ with $\Sigma_2^{(1)}$, say, and 
$x^{7,8}$ with $\Sigma_2^{(2)}$. The circle $x^6$ should 
be identified with the $S^1$. The constant factor $N$ on 
the right-hand-side 
of \dfoatut\ is coming from the D4 branes going around the circle,
while the $r$-dependent piece should be attributed to the  D4 branes
stretching between the NS5 branes. Having these two types of
fourbranes will be important when the solution is lifted to
eleven dimensions.

\subsec{The M-theory versions of KT}
Lifting to eleven-dimensional supergravity will produce two distinct
solutions, corresponding to the two versions of KT in type IIA. In both
cases a new twisted $U(1)$ develops.

Let us start with the second case discussed in  the last section 
 (solution \dsab-\dfoatut) and lift to eleven dimensions.
The solution reads
\eqn\ktlm{\eqalign{
g_s^{2\over 3}ds_{11}^2= ({1\over 9} h^{1\over 2}r^2 )^{1\over 3}
\left( h^{-{1 \over 2}}dx^{\mu} dx_{\mu}+ h^{1\over 2}
(dr^2+{1\over 6} r^2 \sum_{i=1}^2 (d\Sigma_2^{(i)})^2)\right) \cr
+ ({1\over 9} h^{1\over 2}r^2 )^{-{2\over 3}}
\left( dz^2+ g^2_s (d\psi-M\omega_1)^2 \right)
\cr}}
Note that for  $M=0$ the bundle is untwisted and the 
transverse geometry is foliated by 
$S^2\times S^2 \times  S^1\times S^1$.
However in the presence of fractional branes $M \neq 0$
and a twist is induced. The transverse space
becomes foliated by $T^{1,1}\times S^1$.
More accurately, this is a $T^{1,1}$ space whose 
$S^3$ fibre (when $T^{1,1}$ is viewed as $S^3$ over $S^2$)
is replaced by the lens space $S^3/Z_M$.
Indeed, under the redefinition
$\psi \rightarrow -M\psi$,
 $d\psi-M\omega_1$ becomes 
proportional to the 
covariant displacement on the $U(1)/Z_M$ fibre of a $T^{1,1}$
with reversed orientation.  This is reminiscent
of the situation in \dlpb\ (see also \clpc).

We also have
\eqn\ktlf{
F_4=(N+3g_s M^2 ln{r\over r_0})\, \omega_2  \wedge \omega_2
-(3g_s M\,  {dr \over r}\wedge \omega_2+ 
  dA_1 \wedge dz)\wedge (d\psi - M\omega_1)
}
Note that $dF_4=0$ as it should. To prove that, use the fact that 
$\omega_2 \wedge dA_1=0$.

The solution contains M5-brane charge wrapping the $S^2$ of  $T^{1,1}$,
\eqn\mtcb{Q^{(5,1)}=\int_{S^3\times S^1} F_4  }
This will reduce to NS5 wrapping $S^2$ as in \dfab.

Naively, the solution contains M5 branes wrapping both the ``untwisted'' 
$S^1$ and the $U(1)$ fibre of the base $\Sigma^{(1)}_2\times
\Sigma^{(2)}_2$ over the $T^{1,1}$. Their charge 
${\tilde Q}^{(5,2)}$ is given by
\eqn\mtca{{\tilde Q}^{(5,2)}=\int_{\Sigma^{(1)}_2\times \Sigma^{(2)}_2}
F_4 = N+3g_s M^2 ln{r\over r_0}  }
 Since there is no nontrivial $\Sigma^{(1)}_2\times
\Sigma^{(2)}_2$ cycle in $T^{1,1}$ such a ${\tilde Q}^{(5,2)}$ charge can
be taken  to zero,
even though there is non-vanishing $F_4$ flux. 
However upon reduction to IIA, these phantom M5 branes 
come to existence in the form of the D4 branes wrapping the
$S^1$, as in \dfoatut! Also, as noted below \ktlm, in 
the absence of fractional branes the $U(1)$ bundle is untwisted and
again the phantom M5 branes become ordinary M5 branes 
wrapping $S^1\times S^1$. In the following, all the phantom charges
will carry tildes.

This is the first instance in which we encounter a phenomenon whereby there
are fluxes unsupported by the geometry, which however 
 materialize into physical branes  upon untwisting and/or
Hopf reduction. The $U(1)$
(un)twisting  is the key to this, and in a way provides a discrete
analogue of the large N duality between the fluxes and branes. We will
meet more examples of this in the sequel.

Let us compare to the situation in \ghm\ where 
a string with nonzero winding number in the 
background of a Kaluza-Klein monopole was considered. The string 
can unwind since the total space has $\pi_1=0$, but the charge associated 
to the winding number of the string is conserved. From this one
concludes that there is a zero 
mode among the collective 
coordinates of the KK monopole which 
couples to the charge associated with the winding of the string.
In our case $\pi_1= Z_M$ is nonzero, but 
 there are still no nontrivial one-cycles in homology 
($b_1=0$).

 The lifting of  the solution \dsa-\dfoa\ is similar. The resulting
geometry is now $T^{1,-1}\times S^1$ (for $M\neq 0$)
and the charge $Q^{(5,1)} = 0$. 
As before the $S^3$ fibre of the $T^{1,-1}$ is 
replaced by the lens space  $S^3/Z_M$. 
Again, there is a flux through
the 
contractible $\Sigma^{(1)}_2\times \Sigma^{(2)}_2$. 
Upon reduction  this becomes 
the D4-brane flux.

\newsec{The S-dual KT}
We saw in the previous section that upon $T$-dualizing the KT solution 
we get a configuration with ``untwisted'' $\Sigma^{(1)}_2
\times \Sigma^{(2)}_2 \times S^1$ 
geometry for the level surfaces. It would
be interesting to have a situation where by $T$-dualizing 
a IIB solution we get level surfaces with twisted geometry.
The reason why this does not work with the KT solution \ktso\ is 
clear. From the $T$-duality transformations  (A.22)  we see that
in order to get a twisted $U(1)$ fibration upon going from IIB 
to IIA, the original IIB solution would have to have 
$A_1^{NS} \neq 0$. Upon $T$-dualizing this becomes the connection 
$A_1^{(3)}$ of the $U(1)$ fibration in IIA. Since $H_3
=dA_2^{NS}-dA_1^{NS} \wedge dz$, this means that the original IIB
solution would have to have $H_3$ with nonzero flux through the Hopf 
fibre. We can indeed obtain a solution of IIB with the 
aforementioned property by performing an $S$-duality 
transformation on the KT solution:
\eqn\stra{\eqalign{\tau &\rightarrow {a \tau+b\over c\tau+d}\cr
\pmatrix{  H_3 \cr F_3 \cr }&\rightarrow 
\pmatrix{ d & c \cr b & a \cr }\pmatrix{  H_3 \cr F_3 \cr }
\cr}}
with the rest of the fields inert and
\eqn\tbdefs{\tau :=\chi+i e^{-\phi}; \,\,\,\,\,\,\,
\pmatrix{ a & b \cr c & d \cr } \in SL(2, \ZZ)  } 
The KT solution \ktso\ transforms to
\eqn\stkt{\eqalign{
e^{\phi}&=g_s d^2+{c^2\over g_s}:={\tilde g}_s \cr
\chi&={ac+bd g^2_s\over c^2+d^2 g^2_s}\cr
 H_3&=3d\, g_s M\, {dr\over r}\wedge \omega_2 +c\, M \omega_3  \cr
 F_3&=3b\, g_s M\, {dr\over r}\wedge \omega_2+ a\, M \omega_3
\cr}}
The metric and the five-form field strength are still given by 
\ktso-\ktcf. Reducing to nine dimensions $T$-dualizing and lifting to IIA
we get  two possible situations depending on the choice of relative
orientation of the two $S^2$. Namely, 
a geometry with level surfaces given either by $T^{1,1}$ 
with reversed orientation or  by $T^{1,-1}$.
\eqn\sktm{
\eqalign{
{ g}_s^{1 \over 2}ds_{10}^2=({{\tilde g}_s\over g_s})^{1\over 8}
 ({1\over 9} h^{1\over 2}r^2 )^{1\over 4}
\left( h^{-{1 \over 2}}dx^{\mu} dx_{\mu}+ h^{1\over 2}
(dr^2+{1\over 6} r^2 \sum_{i=1}^2 (d\Sigma_2^{(i)})^2) \right) \cr
+({{\tilde g}_s\over g_s})^{-{7\over 8}}({1\over 9} h^{1\over 2}r^2 )^{-{3\over 4}}  
(dz-c\, M \omega_1)^2
\cr}
}
Note that $dz-c \, M\omega_1$ is 
proportional to the covariant displacement on the $U(1)/Z_{c\, M}$
fibre either of a $T^{1,1}$ with {\it reversed} orientation
or of a $T^{1,-1}$.
We have
\eqn\tivp{\eqalign{
F_2&=(a-\chi c)\, M \omega_2 \cr
F_3&=3d\, g_s  M{dr\over r}\wedge \omega_2
+d A_1 \wedge (dz-c \, M\omega_1)\cr
F_4&=(N+3g_sM^2 ln{r\over r_0}) \omega_2\wedge \omega_2
+3(b-\chi d)\, M\, g_s{dr\over r}\wedge \omega_2 \wedge
 (dz-c \, M\omega_1)\cr
}}
The brane content is different in each case.

First we discuss the $T^{1,1}$ with reversed orientation.
The solution has zero units of D6 branes wrapping 
$S^3$,
\eqn\gjhy{ Q^{(6)}= \int_{S^2} F_2 =0 }
There is however nonzero net $Q^{(5)}$ charge corresponding
to NS5 branes wrapping $S^2$,
\eqn\dfatr{
Q^{(5)}=\int_{S^3} F_3 = -cM
}
Finally, there is a ${\tilde Q}^{(4)}$ phantom charge corresponding to D4
branes wrapping the Hopf fibre,
\eqn\dfoatr{
{\tilde Q}^{(4)}=\int_{\Sigma^{(1)}_2\times \Sigma^{(2)}_2} F_4 
= N+3g_s M^2 ln{r\over r_0}
}
Again, as in \mtca, note that this makes sense either
upon Hopf reducing or in the absence of fractional branes 
in which case the $U(1)$ fibre gets untwisted.

Turning to the $T^{1,-1}$ case we find $Q^{(6)}=(a-\chi c)\, M$ 
and $Q^{(5)}=0$. 
What
was said  before about \dfoatr\ is true here as well.

\subsec{The M-theory lift}

Upon lifting to M-theory, yet another $U(1)$ fibre develops. 
The resulting transverse geometry is foliated by an 
$S^1/Z_{aM}$ bundle over $T^{1,1}/Z_{cM}$ 
such that reducing the fibre 
along any one of the two circles produces a $T^{1,1}$ space
(or a $T^{(1,-1)}$, by a reasoning that should be familiar by now).
In this sense there is a similarity with the AMV solution in the 
next section.\foot{In spite of the similarity it is easy to see that
the resulting geometry cannot be that of AMV simply because the original
ten-dimensional one is not that of the resolved conifold.} 
The transverse geometry in
that case is an $S^3$ over
 $S^3$ bundle. Viewing the base as a Hopf fibration over $S^2$ and reducing
along the $U(1)$ fibre produces an $S^3$ (up to moding out by a 
discrete group) over $S^2$ bundle, which
is actually a $T^{1,1}$ space.

Although the geometry appears to be complicated, it is in fact related 
by a simple coordinate transformation to the geometry of the solution 
in section 3.2.

Analytically, the eleven-dimensional metric reads 
\eqn\ktlmsdf{\eqalign{
g_s^{{2\over 3}} ds_{11}^2= ({1\over 9} h^{1\over 2}r^2 )^{1\over 3}
\left( h^{-{1 \over 2}}dx^{\mu} dx_{\mu}+ h^{1\over 2}
(dr^2+{1\over 6} r^2 \sum_{i=1}^2 (d\Sigma_2^{(i)})^2)\right) \cr
+{g_s\over \tilde{g}_s} ({1\over 9} h^{1\over 2}r^2 )^{-{2\over 3}}
\left( (dz-c\, M\omega_1)^2
+\tilde{g}^2_s[(dw-a\, M \omega_1)-\chi(dz-cM\omega_1)]^2 \right)
\cr}}
while the four-form is given by 
\eqn\mfhkj{\hat{F_4}=F_4-F_3\wedge [(dw-a\, M \omega_1)
-\chi(dz-c\, M\omega_1)]}
with $F_4, F_3$ as in \tivp.

It is easy to check that under the coordinate transformation
\eqn\coordtr{\pmatrix{ w \cr 
z \cr }\rightarrow \pmatrix{ d &-b \cr 
c & -a \cr }\pmatrix{ w \cr 
z\cr }    }
the metric reduces to \ktlm! 

The solution contains phantom M5 branes wrapping 
the two circles. 
Their charge  ${\tilde Q}^{(5,1)}$ is given by
\eqn\mtcaoi{{\tilde Q}^{(5,1)}=\int_{\Sigma^{(1)}_2\times \Sigma^{(2)}_2}
F_4 = N+3g_s M^2 ln{r\over r_0}  }
Upon reduction to IIA these become the D4 branes wrapping the
Hopf fibre of $T^{1,1}$, as in \dfoatr.
Depending on the orientation there can be phantom
M5-brane charge wrapping  $S^2$,
\eqn\mtcboiu{{\tilde Q}^{(5,2)}=\int_{S^1\times S^3} F_4 }
which reduces to NS5 wrapping $S^2$ as in \dfatr.

If the opposite orientation is chosen, $Q^{(5,2)}=0$ which of course 
reduces to $Q^{(5)}=0$ in IIA.

\newsec{Fivebranes in the AMV Solution}

The AMV solution \amv\ involving M-theory on a manifold of $G_2$ holonomy
(locally an $S^3 \times \IR^4$) reduced to 
ten dimensions is given by warped 4d
Minkowski times the resolved conifold of \refs{\cdlo, \zt}. In addition,
there is a nonzero $F_2$ flux through the $S^2$.  As before, modulo the
large N duality ($S^3$ flop), we think of this as the supergravity
solution corresponding to D6 wrapping the $S^3$ with a stabilizing flux
through $S^2$ and thus producing the geometry of the resolved conifold.
For some related work see also \refs{\mnz, \DEdelsteinNunez, 
\DasguptaOhTatar,
\cglpa, 
\AganagicVafa, 
\cglpb, 
\BrandhuberGomisSGubserGukov}. 

From (A.22) we immediately see that upon $T$-dualizing 
this solution to IIB, we get ``untwisted'' $\Sigma^{(1)}_2
\times \Sigma^{(2)}_2 \times S^1$ geometry: 
In order to get a twisted $U(1)$ fibration upon going from IIA to
IIB, the original IIA solution would have to have $A_1^{(2)} \neq
0$. Upon $T$-dualizing this becomes the connection $A_1$ of the $U(1)$
fibration in IIB.  Since, as we see from  (A.14),
$\hat{F}_3=F_3^{(1)}-dA_1^{(2)} \wedge e^5$, this means that the
original IIA solution would have to have an $F_3$ with nonzero flux
through the Hopf fibre. This would signal the presence of NS5 branes
wrapping the $S^2$, which are absent from the solution of \amv.

This brings us to the question of whether such a configuration of NS5's
and D6's is known and if so, what would it lift to in M-theory. Note that
the fivebrane and sixbranes stabilize the $S^2$ and $S^3$ factors
respectively, The $S^2$ which is wrapped by the NS5 upon lifting becomes
the base of a twisted $U(1)$ fibration and thus is no longer  able
to support an M5. This is in agreement with the fact that $G_2$ holonomy
manifolds do not have calibrated two-dimensional submanifolds.

We should therefore look for an eleven-dimensional $G_4$ that has a
flux through the (eleven-dimensional) Hopf fibre. Reducing this will
give an $F_3$ flux, which upon integration over the $S^3$ 
will give NS5 charge. Since there are two $S^3$'s in the original
geometry, an obvious possibility is to consider a flux
\eqn\gflux{\int_{S^3 \times S1} G_4 }
which upon  reduction will give an NS5 wrapping $S^2$. Although the
ten-dimensional geometry is foliated by $T^{1,1}$, it  cannot be that
of the resolved conifold, since the flux of the NS three-form will
stabilize the $S^3$ part as well. At the moment we are not able to write
down explicitly such a solution containing simultaneously D6 branes
wrapping $S^3$ and NS5 branes wrapping $S^2$ and having four common
non-compact directions. 

M-theory solutions with a flux as in \gflux\ have recently 
appeared in the literature \refs{\cglpa, \cglpb}. 
It would clearly
be interesting to obtain their IIA reductions and their IIB $T$-duals. 
The hope is to provide 
 an explicit realization of the mirror symmetry
between D5 branes on $S^2$ of the deformed conifold in type IIB
and D6 on $S^3$ of 
the resolved in type IIA,  
as a single $T$-duality along the Hopf fibre of the $T^{1,1}$.\foot{Note
that as we have already remarked, the existence
of NS5 branes is required in addition to the D branes
for this duality to work.}
This task is complicated, however, by the form of the deformed conifold 
metric, as will be discussed in the next section.

\newsec{The KS Solution}

Although we have already excluded the possibility that the KS solution
\ks\ is the $T$-dual of the AMV solution --since
$T$-dualizing the latter
gives untwisted transverse geometry whereas the former has a twisted
one-- we can still try to perform the analysis of the previous
sections to this case. 
Technically this is complicated by the fact that the metric of 
the deformed conifold 
is written in terms of variables in which the $U(1)$ isometry 
is implicit \mt. Here we make partial progress by identifying a set of new 
variables in which the $U(1)$ isometry of the metric will be manifest,
but we are yet unable to write down explicitly the metric 
in terms of these new variables.

The KS solution reads
\eqn\kss{\eqalign{  
ds^2_{10}&=h^{-1/2}(\tau)dx_{\mu}dx^{\mu}+h^{1/2}(\tau)ds_6^2\cr
F_3&=M(e^5\wedge g^3\wedge g^4+d[F(\tau)(g^1\wedge g^3+g^2\wedge g^4)])\cr
B_2&=g_s M[f(\tau)g^1\wedge g^2+k(\tau)g^3\wedge g^4]\cr
F_5&=g_s M^2 l(\tau)g^1\wedge g^2\wedge g^3\wedge g^4\wedge e^5,
\cr}}
where
\eqn\dcmiks{\eqalign{ ds_6^2={1\over 2}\epsilon^{4/3}K(\tau)
( {1\over 3K(\tau)}
(d\tau^2+(e^5)^2)+cosh^2({\tau \over 2})[(g^3)^2+ (g^4)^2]\cr
+sinh^2({\tau \over 2})[(g^1)^2+ (g^2)^2] )
\cr}}
and $\tau$ is a radial coordinate. 
The explicit form of the functions $F,f,k,l,K$ 
will not be important in the following. The one-forms $g^i$ are 
defined as
\eqn\basa{\eqalign{
g^1={e^1-e^3 \over \sqrt{2}}&; \,\,\,\,\,\,\,
g^2={e^2-e^4 \over \sqrt{2}}\cr
g^3={e^1+e^3 \over \sqrt{2}}&; \,\,\,\,\,\,\,
g^4={e^2+e^4 \over \sqrt{2}}\cr
g^5&=e^5 \cr}}
so that the level surfaces of \dcmiks\ are of the general form \tb.

As we noted in section 4, for this solution to $T$-dualize to
a twisted transverse geometry, 
$B_2$ would have to have a nonzero component along
the direction corresponding to the $U(1)$ isometry.

Trying to Hopf-reduce runs into trouble because 
the solution is written in terms of coordinates in which 
the $U(1)$ isometry is not simply a shift in $\psi$. In fact
from \baseb, \tb\ we see that the isometry reads
\eqn\isom{\eqalign{\psi &\rightarrow \psi+c; \cr
\pmatrix{  sin\theta_2 d\phi_2\cr d\theta_2\cr } &\rightarrow  
\pmatrix{ cos c &
sin c \cr  -sin c &  cos c \cr }
\pmatrix{  sin\theta_2 d\phi_2\cr d\theta_2\cr }
\cr}}
We would like to find a coordinate transformation
for $\theta_2, \phi_2$ 
which has the effect of the second line in \isom.
Clearly this would have to be a special case of the $SU(2)$
group of isometries of the sphere
\eqn\sphis{z\rightarrow {a z+ b\over -\bar{b}z+\bar{a}};
\,\,\,\,\,\,\, z:=e^{i\phi_2}tan{\theta_2\over 2};  
\,\,\,\,\,\,\, |a|^2+|b|^2=1.  }
Indeed for $a=cos{\e \over 2}, \,\, b=sin{\e \over 2}$, 
\sphis\ implies \isom\ to order ${\cal O}(\e)$, provided we identify
\eqn\cid{c=\e {sin\p_2 \over sin\th_2}}
Explicitly the coordinate transformation  reads, 
to ${\cal O}(\e)$ order,
\eqn\expsph{ \eqalign{
\psi &\rightarrow \psi+\e {sin\p_2 \over sin\th_2}\cr
\phi &\rightarrow \phi -\e sin\phi_2 cot\th_2\cr
\th_2 &\rightarrow \th_2 +\e cos\phi_2
\cr}}
It is easy to check that \expsph\ leaves $e^5$ invariant.
We have therefore succeeded in constructing the killing vector $k$
corresponding to the $U(1)$ isometry,
\eqn\kill{k={sin\p_2 \over sin\th_2} {\partial \over \partial \psi}
-sin\phi_2 cot\th_2 {\partial \over \partial \phi_2}
+cos\phi_2 {\partial \over \partial \th_2}}
It remains to find a set of new variables $\psi', \th_2', \phi_2'$
in which $k$ locally takes the form $k=\partial / \partial \psi'$. 
Rewriting \dcmiks\ in terms of these new variables would result in
a metric whose components do not depend on $\psi'$ and Hopf reduction
would proceed as before.
The variables we are looking for would therefore 
have to satisfy the equation
\eqn\killtrnsf{\pmatrix{ {\partial \theta_2' \over \partial \theta_2} 
 & {\partial \theta_2' \over \partial \phi_2} & 
{\partial \theta_2' \over \partial \psi}\cr
{\partial \phi_2' \over \partial \theta_2} 
 & {\partial \phi_2' \over \partial \phi_2} & 
{\partial \phi_2' \over \partial \psi}\cr
{\partial \psi' \over \partial \theta_2} 
 & {\partial \psi' \over \partial \phi_2} & 
{\partial \psi' \over \partial \psi}\cr }
\pmatrix{k^{\th_2}\cr k^{\phi_2} \cr k^{\psi}\cr}=
\pmatrix{0\cr 0\cr 1\cr}}
One can easily verify that
\eqn\tipipi{\eqalign{
\hat{L}{1\over 2}(f_1+f_2)&={sin\phi_2 \over sin\th_2};
\,\,\,\,\,\,\,\, \hat{L}{1\over 2}(f_1-f_2)=1 \cr
&\hat{L}h(sin\phi_2 sin\th_2)=0
\cr}}
where
\eqn\eqnasde{\eqalign{\hat{L}&:=k-k^{\psi}{\partial \over \partial \psi};\cr
f_{1,2}(\phi_2,\th_2)&:= arctan 
\left( {sin\phi_2 \pm sin\th_2\over cos\phi_2 cos\th_2} \right) \cr 
}}
and $h$ is an arbitrary function of $sin\phi_2 sin\th_2$.

Taking \tipipi\ into account, we can write down an explicit 
special solution to \killtrnsf:\foot{This is not the most general
form. One can add arbitrary functions of $ sin\phi_2 sin\th_2$ to the rhs.
Also one could take arbitrary linear combinations of $\th_2', \,
\phi_2'$ in the second and third lines.}
\eqn\exss{\eqalign{\psi' &=  \psi + f_1(\phi_2,  \th_2) \cr
\phi' &=  \psi + {1\over 2} \left( f_1(\phi_2,  \th_2)
+f_2(\phi_2,  \th_2) \right)\cr
\th'&= h(sin\phi_2 sin\th_2)
\cr}}
The Jacobian of the above transformation reads
\eqn\jac{J= sin\th_2 h'(x)\vert_{x=sin\phi_2 sin\th_2} }
Finally one can derive the relation of the differentials of the old
coordinates in terms of the new,
\eqn\rlotd{\pmatrix{d\phi_2\cr
d\th_2 \cr
d\psi\cr}=
\pmatrix{-cot\th_2 sin\phi_2 & cot\th_2 sin\phi_2 & {cos\phi_2\over
h'(cos^2\phi_2+cos^2\th_2sin^2\phi_2)sin\th_2 }\cr
cos\phi_2 & -cos\phi_2 & {cos\th_2 sin\phi_2\over
h'(cos^2\phi_2+cos^2\th_2sin^2\phi_2) }\cr
{sin\phi_2 \over sin\th_2} & 1-{sin\phi_2 \over sin\th_2} &
 -{cos\phi_2 cot\th_2\over
h'(cos^2\phi_2+cos^2\th_2sin^2\phi_2) }\cr  }    
\pmatrix{d\psi'\cr d\phi_2'\cr d\th_2'\cr}
}
Using the above relation, one can examine whether or not $B_2$ in \kss\
has a non-trivial component along the $U(1)$ fibre. The answer is that 
it does and therefore, as already explained, the $T$-dual version of KS 
has twisted transverse geometry.

\bigskip
\centerline{\bf Acknowledgments}\nobreak
\bigskip

\noindent It is a pleasure to thank C. Bachas, M. Cvetic, M. Douglas and 
G. Ferretti 
for discussions. We would also like to
thank K. Becker and S. Corley for pointing out typos.
The work of RM  is supported in part by EU contract
HPRN-CT-2000-00122 and by INTAS contract 55-1-590; DT is supported in part
by EU contract HPRN-CT-2000-00122 and by the SNSRC. DT would like to
thank Ecole Normale Sup{\'e}rieure and CERN for hospitality.

\bigskip
\noindent {\bf Note added:} After this paper was posted on the hep-th
archive, we were informed by I. Klebanov, A. Tseytlin and L. Pando Zayas of
their related independent work which overlaps with section 3 of this
paper. See the talk by I. Klebanov at the Avatars of M-theory conference
(UCSB) {\tt http://online.itp.ucsb.edu/online/mtheory\_c01/klebanov/}.

\vfil\eject

\appendix{A}{Review of KK on a circle.}

This is a collection of useful formulae. There is nothing here that cannot
be found in the literature. However we include this appendix as
a self-contained set of conventions.

\subsec{Reduction of 11d supergravity to 10d type IIA}

The starting point is the eleven-dimensional Lagrangian
\eqn\mthe{
\hat{e}^{-1}L=\hat{R} -{1 \over 48} \hat{F}_4^2 +C.S.  
}
where 
\eqn\mfst{
\hat{F}_4=d\hat{A}_3
}
Note that all eleven- (ten-) dimensional
field strengths and potentials 
are denoted with (without) a hat.
Assuming the geometry has a $U(1)$ isometry, we can 
put the metric in the form:
\eqn\mmet{
ds_{11}^2=e^{-{1\over 6}\varphi}ds_{10}^2+ 
e^{{4\over 3}\varphi}(A_1+dz)^2  
}
where  $\varphi$, $A_1$, $\hat{A}_3$ and the metric in $ds_{10}^2$, are 
all assumed independent of the $U(1)$ coordinate $z$. 
We also reduce the 3-form potential
\eqn\mpot{
\hat{A}_3=A_3+A_2 \wedge dz
}
The eleven-dimensional Lagrangian becomes
\eqn\redm{
e^{-1}L= R  -{1\over 2} (\partial \varphi)^2
-{1 \over 4} e^{{3\over 2}\varphi} F_2^2
-{1 \over 12}e^{-\varphi} F_3^2
-{1 \over 48}e^{{1\over 2}\varphi} F_4^2 +C.S.   
}
where:
\eqn\redf{
F_4=dA_3+dA_2 \wedge A_1; \,\,\,\,\, F_3=-dA_2; \,\,\,\,\, F_2=dA_1
}
so that
\eqn\mrfs{
\hat{F}_4= F_4+F_3 \wedge (dz+A_1)
}

\subsec{Reduction of 10d type IIA to 9d N=2}

We start with the Lagrangian 
\eqn\tala{
\hat{e}^{-1}L=\hat{ R}  -{1\over 2} (\partial \hat{\phi})^2
-{1 \over 4} e^{{3\over 2}\hat{\phi}} \hat{F}_2^2
-{1 \over 12}e^{-\hat{\phi}} \hat{F}_3^2
-{1 \over 48}e^{{1\over 2}\hat{\phi}} \hat{F}_4^2 +C.S.   
}
where
\eqn\tafs{
\hat{F}_4=d\hat{A}_3+d\hat{A}_2 \wedge \hat{A}_1;
\,\,\,\,\, \hat{F}_3=d\hat{A}_2; \,\,\,\,\, \hat{F}_2=d\hat{A}_1
}
We assume that there is a $U(1)$ isometry so that the ten-dimensional 
metric can be cast in the form
\eqn\tame{
ds_{10}^2=e^{-{1\over 2\sqrt{7}}\varphi}ds_{9}^2+ 
e^{{\sqrt{7} \over 2}\varphi}(A^{(3)}_1+dz)^2 
}
where $\varphi$, $A^{(3)}_1$ and the nine-dimensional metric $ds_{9}^2$ 
do not depend on the 
$U(1)$ coordinate $z$.
We reduce the potentials as follows
\eqn\tapo{
\hat{A}_1=A_1^{(1)}+A_0  dz; \,\,\,\,\,
\hat{A}_2=A_2^{(1)}+A_1^{(2)} \wedge dz; \,\,\,\,\,
\hat{A}_3=A_3+A^{(2)}_2 \wedge dz
}
The Lagrangian reduces to
\eqn\tarl{
\eqalign{
e^{-1}L= R  &-{1\over 2} (\partial \varphi)^2
-{1\over 2} (\partial \phi)^2
-{1 \over 2} e^{{3\over 2}\phi-{\sqrt{7}\over 2}\varphi} (F_1)^2\cr
&-{1 \over 4} e^{-\phi-{3\over \sqrt{7}}\varphi} (F_2^{(2)})^2
-{1 \over 4} e^{{3\over 2}\phi+{1\over 2\sqrt{7}}\varphi} (F_2^{(1)})^2 
-{1 \over 4}e^{{4\over \sqrt{7}}\varphi} (F_2^{(3)})^2\cr
&-{1 \over 12} e^{-\phi+{1 \over \sqrt{7}}\varphi} (F_3^{(1)})^2
-{1 \over 12} e^{{1\over 2}\phi-{5\over 2\sqrt{7}}\varphi} (F_3^{(2)})^2
-{1 \over 48}e^{{1\over 2}\phi+{3\over 2\sqrt{7}}\varphi } (F_4)^2 +C.S.\cr
} 
}
where
\eqn\tarf{
\eqalign{
\phi=\hat{\phi};& \,\,\,\,\,\,\, F_1=-dA_0 \cr
F_2^{(1)}=dA_1^{(1)}+dA_0 \wedge A_1^{(3)}; &\,\,\,\,\,\,\,
F_2^{(2)}=-dA_1^{(2)}; \,\,\,\,\,\,\,
F_2^{(3)}=dA_1^{(3)}\cr
F_3^{(1)}=dA_2^{(1)}+dA_1^{(2)}\wedge A_1^{(3)}; \,\,\,\,\,\,\,
F_3^{(2)}&=-dA_2^{(2)}+A_1^{(1)}\wedge dA_1^{(2)}+A_0 dA_2^{(1)} \cr
F_4=dA_3+dA_2^{(1)}\wedge A_1^{(1)}
+dA_2^{(2)}\wedge A_1^{(3)}&-dA_1^{(2)}\wedge A_1^{(1)}\wedge A_1^{(3)}
-A_0 dA_2^{(1)}\wedge A_1^{(3)}  \cr
}
}
so that
\eqn\bpourdoumpas{\eqalign{
\hat{F}_4&=F_4+F_3^{(2)} \wedge (dz+A_1^{(3)})\cr
\hat{F}_3&=F_3^{(1)}+F_2^{(2)} \wedge (dz+A_1^{(3)})\cr
\hat{F}_2&=F_2^{(1)}+F_1 \wedge (dz+A_1^{(3)})\cr
}}

\subsec{Reduction of 10d type IIB to 9d N=2.}

We start with the Lagrangian 
\eqn\tbla{
\hat{e}^{-1}L=\hat{ R} -{1\over 2} (\partial \hat{\phi})^2
-{1 \over 2} e^{2\hat{\phi}} (\partial \hat{\chi})^2
-{1 \over 12}e^{\hat{\phi}} (\hat{F}_3)^2
-{1 \over 12}e^{-\hat{\phi}} (\hat{H}_3)^2
-{1 \over 4\times 5!} (\hat{F}_5)^2 +C.S. 
}
where
\eqn\tbfs{
\eqalign{
\hat{F}_5=d\hat{A}_4^R-{1\over 2}  d\hat{A}_2^{NS} \wedge \hat{A}_2^R
+{1\over 2} d \hat{A}_2^R \wedge \hat{A}_2^{NS}= {}^*\hat{F}_5 \cr
\hat{F}_3=d\hat{A}_2^R-\hat{\chi} d\hat{A}_2^{NS} ; 
\,\,\,\,\, \hat{H}_3=d\hat{A}_2^{NS}\cr
}
}
We assume that there is a $U(1)$ isometry so that the ten-dimensional 
metric can be cast in the form
\eqn\tbme{
ds_{10}^2=e^{-{1\over 2\sqrt{7}}\varphi}ds_{9}^2+ 
e^{{\sqrt{7} \over 2}\varphi}(A_1+dz)^2 
}
where $\varphi$, $A_1$ and the nine-dimensional metric $ds_{9}^2$ 
do not depend on the 
$U(1)$ coordinate $z$.
We reduce the potentials as follows
\eqn\tbpo{\eqalign{
\hat{A}_2^R=A_2^R+A_1^R \wedge dz; \,\,\,\,\,\,\,
&\hat{A}_2^{NS}=A_2^{NS}+A_1^{NS} \wedge dz \cr
\hat{A}_4^R=&A_4^R+A_3^R \wedge dz\cr
}}
The Lagrangian reduces to
\eqn\tbrl{
\eqalign{
e^{-1}L= R  &-{1\over 2} (\partial \varphi)^2
-{1\over 2} (\partial \phi)^2
-{1 \over 2} e^{2\phi} (\partial \chi)^2\cr
&-{1 \over 4} e^{-\phi-{3\over \sqrt{7}}\varphi} (F_2^{NS})^2
-{1 \over 4} e^{\phi-{3\over \sqrt{7}}\varphi} (F_2^R)^2
-{1 \over 4}e^{{4\over \sqrt{7}}\varphi} (F_2)^2\cr
&-{1 \over 12} e^{-\phi+{1 \over \sqrt{7}}\varphi} (F_3^{NS})^2
-{1 \over 12} e^{\phi+{1\over \sqrt{7}}\varphi} (F_3^R)^2
-{1 \over 48}e^{-{2\over \sqrt{7}}\varphi } (F_4^R)^2 +C.S.\cr
} 
}
where
\eqn\tbfs{
\eqalign{
\phi=\hat{\phi};& \,\,\,\,\,\,\, \chi=\hat{\chi}\cr
F_2^{R}=-dA_1^{R}+\chi d A_1^{NS};& \,\,\,\,\,\,\,
F_2^{NS}=-dA_1^{NS}; \,\,\,\,\,\,\,
F_2=dA_1\cr
F_3^{NS}=dA_2^{NS}+A_1 \wedge dA_1^{NS};  \,\,\,\,\,\,\,
F_3^R=dA_2^{R}&+dA_1^{R}\wedge A_1
-\chi(dA_2^{NS}+dA_1^{NS}\wedge A_1)\cr
F_4^R=-dA^R_3+{1\over 2}(- dA_2^{NS}\wedge A_1^R
-A_2^{NS}&\wedge dA_1^R
+dA_2^R\wedge A_1^{NS}+A_2^R\wedge dA_1^{NS})\cr
}
}
so that

\eqn\taff{
\hat{F}_n=F_n+F_{n-1} \wedge (dz+A_1)
}

\subsec{9d T-duality}

The two nine-dimensional Lagrangians described above, are in fact related
to each other by the following local field transformations:
\eqn\tdtr{
\eqalign{
\phi_A \leftrightarrow {3\over 4} \phi_B-{\sqrt{7}\over 4} \varphi_B; 
\,\,\,\,\,\,\,
\varphi_A \leftrightarrow -{\sqrt{7}\over 4} \phi_B-{3\over 4} \varphi_B;
 \,\,\,\,\,\,\, A_0 \leftrightarrow -\chi \cr
A_1^{(1)}\leftrightarrow  -A_1^{R}+\chi A_1^{NS}; \,\,\,\,\,\,\,
A_1^{(2)}\leftrightarrow  -A_1; \,\,\,\,\,\,\,
A_1^{(3)}\leftrightarrow  -A_1^{NS} \cr
A_2^{(1)} \leftrightarrow A_2^{NS}-A_1^{NS}\wedge A_1; \,\,\,\,\,\,\,
A_2^{(2)}\leftrightarrow  -A_2^{R}+A_1^{R}\wedge A_1\cr
A_3\leftrightarrow  -A_3^R+{1\over 2} A_1^{NS}\wedge A_2^{R}
-{1\over 2} A_1^{R}\wedge A_2^{NS}
-A_1^{NS}\wedge A_1^{R}\wedge A_1 \cr
}
}
Or, in terms of field strenghts:
\eqn\tdtrfss{
\eqalign{
F_1 &\leftrightarrow d\chi \cr
F_2^{(1)} \leftrightarrow F_2^R;\,\,\,\,\,\,\,
F_2^{(2)} &\leftrightarrow F_2; \,\,\,\,\,\,\,
F_2^{(3)} \leftrightarrow F_2^{NS}\cr
F_3^{(1)} \leftrightarrow F_3^{NS};&  \,\,\,\,\,\,\,
F_3^{(2)} \leftrightarrow F_3^{R} \cr
F_4 &\leftrightarrow F_4^R; \cr
}}

\subsec{String vs Einstein frame.}

In ten dimensions the Einstein metric is related to the string metric
through
\eqn\seme{
g_{\mu \nu}=e^{-{1\over 2}\hat{\phi}}g^{str}_{\mu \nu}
}
The reduction of the ten-dimensional metric in the string frame reads
\eqn\sere{
ds_{10}^{str}=e^{{1\over 2}(\phi-{1\over \sqrt{7}}\varphi)}ds_9^2
+e^{{1\over 2}(\phi+ \sqrt{7}\varphi)}(dz+A_1)^2
}
In the string frame the ten-dimensional type IIB
Lagrangian becomes
\eqn\sefi{
\hat{e}^{-1}L= e^{-2\hat{\phi}}(\hat{ R} 
+4 (\partial \hat{\phi})^2
-{1 \over 12} (\hat{H}_3)^2)
-{1 \over 2} (\partial \hat{\chi})^2
-{1 \over 12} (\hat{F}_3)^2
-{1 \over 4\times 5!} (\hat{F}_5)^2 +C.S.   
}

\listrefs
\end